\documentclass[runningheads]{svmult}

\usepackage{makeidx}   
\usepackage{graphicx}  
\usepackage{subeqnar}  
\usepackage{multicol}  
\usepackage{physprbb}  

\usepackage{astrobib}
\usepackage{journals}

\makeindex             

\begin{document}
\title*{The power of jets: New clues from radio circular polarization and X-rays}
\toctitle{The power of jets: New clues from radio circular polarization and X-rays}
%
%
\titlerunning{Jets: Circular Polarization and X-rays}
%
\author{Heino Falcke\inst{1}
\and Thomas Beckert\inst{1}
\and Sera Markoff \inst{1}
\and Elmar K\"ording\inst{1}
\and Geoffrey C. Bower\inst{2}
\and Rob Fender\inst{3}}
\authorrunning{Falcke et al.}
%
%
\institute{Max-Planck-Institut f\"ur Radioastronomie, Auf dem H\"ugel 69, D-53121 Bonn, Germany
\and UC Berkeley, RAL, 601 Campbell Hall, Berkeley CA 94720, USA 
\and Astronomical Institute "Anton Pannekoek"and Center for High Energy Astrophysics, University of Amsterdam, Kruislaan 403, 1098 SJ Amsterdam, The Netherlands}

\maketitle              

\begin{center}{\it to appear in: ``Lighthouses of the Universe'', Springer
Verlag, ESO Astrophysics Symposia, Eds: R.Sunyaev, M.Gilfanov,
E.Churazov}
\end{center}

\begin{abstract}
Jets are ubiquitous in accreting black holes. Often ignored, they may
be a major contributor to the emitted spectral energy distribution for
sub-Eddington black holes. For example, recent observations of
radio-to-X-ray correlations and broad band spectra of X-ray binaries
in the low/hard state can be explained by a significant synchrotron
contribution from jets also to their IR-to-X-ray spectrum as proposed
by \citeN{MarkoffFalckeFender2001}. This model can also explain
state-transitions from low/hard to high/soft states. Relativistic
beaming of the jet X-ray emission could lead to the appearance of
seemingly Super-Eddington X-rays sources in other galaxies. We show
that a simple population synthesis model of X-ray binaries with
relativistic beaming can well explain the currently found distribution
of off-nucleus X-ray point sources in nearby galaxies. Specifically we
suggest that the so-called ultra-luminous X-ray sources (ULXs, also
IXOs) could well be relativistically beamed microblazars. The same
model that can be used to explain X-ray binaries also fits
Low-Luminosity AGN (LLAGN) and especially Sgr A* in the Galactic
Center. The recent detection of significant circular polarization in
AGN radio cores, ranging from bright quasars down to low-luminosity
AGN like M81*, Sgr A* and even X-ray binaries, now places additional
new constraints on the matter contents of such jets. The emerging
picture are powerful jets with a mix of hot and cold matter, a net
magnetic flux, and a stable magnetic north pole.
\end{abstract}

\section{Introduction}
Quasars are probably the most important lighthouses of the universe,
because they are extremely luminous, compact, and emit at all
observable wavelengths. Most of the radio and high energy emission in
quasars can be attributed to relativistic jets produced in the
vicinity of the central black hole. In blazars also infrared, optical
and X-ray emission is produced by these jets. Despite their prominence
at all these energies, some basic properties of jets have not been
clarified despite intense observing campaigns over the last three
decades. Some of these questions are: Why and how are jets produced?
What matter are they made of? How much energy is carried in the jet?
Why are some jets more radio-loud than others? What is the exact
magnetic field configuration?

A possible explanation for the slow progress in answering some of
these questions could be that, in contrast to many other astrophysical
phenomena, jets were first discovered at rather large distances,
i.e. in distant quasars and only in a few cases closer to home. The
reason for this is most likely that the radio luminosity of jets
scales non-linearly with jet power and is in fact relatively weaker in
low-luminosity jets than in quasars (see for example
\citeNP{FalckeBiermann1995}). Only in recent years have we become to
appreciate that relativistic jets exist over a wide range of distances
and over a wide range of black hole masses and accretion rates. This
now opens up the possibility to study the physics of jets in a much
larger parameter range and to revisit some of the early questions in
this context. Here we will concentrate on two relatively new ways of
approaching jet physics, namely through their X-ray emission and their
circular polarization.

\section{X-rays from Microquasar Jets}
One important finding of recent years was that X-ray binaries (XRBs)
possess relativistic jets as well \cite{MirabelRodriguez1994}, leading
to the term ``microquasar''.  So far these jets have mainly been seen
in radio emission, but we will argue here that emission in other
wavebands, specifically NIR and X-rays, is almost unavoidable.

\begin{figure}
\begin{center}
\includegraphics[width=\textwidth]{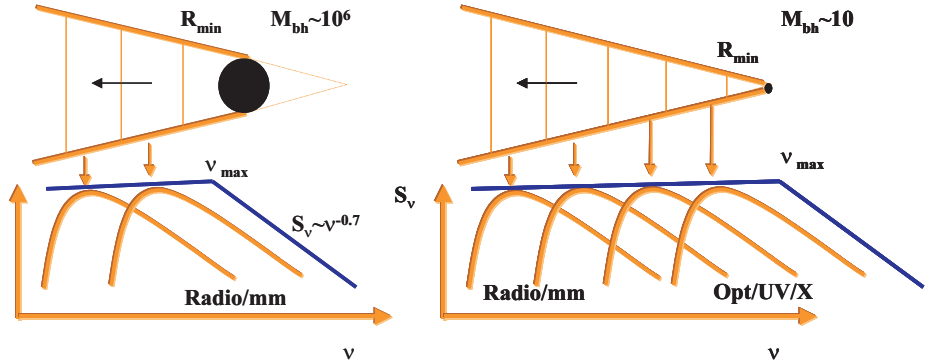}
\end{center}
\caption[]{In a self-similar conical jet model the frequency is inversely
proportional to the size. The turn-over frequency of the flat,
optically thick part of the jet spectrum is determined by the smallest
scale in the system, which should scale with the gravitational radius
and the mass of the black hole. Hence, the turnover which in Sgr A*
(M$\sim10^6M_\odot$) or a quasar occurs somewhere in the submm-range
should shift into the optical/X-ray regime for a stellar mass black
hole.}
\label{falcke-f1}
\end{figure}

A characteristic feature of jets, also in X-ray binaries, is their
flat-spectrum radio core which is best seen during phases of relative
quiescence (e.g. as in GRS 1915+105,
\cite{DhawanMirabelRodrguez2000}). \citeN{Fender2001} found that the
low/hard-state of the X-ray spectrum is correlated with the presence
of a persistent flat-spectrum radio core. He also argued that this
flat spectrum of the synchrotron emitting radio core extends up into
the near-infrared and perhaps optical regime. This is, in fact, not
surprising. The standard model for flat radio spectra
\cite{BlandfordKonigl1979,HjellmingJohnston1988,FalckeBiermann1999}
suggests that the emission arises from self-absorbed sections in a
conical jet, where the smallest scales contribute at the highest
frequencies (size $\propto$ $\nu^{-1}$). This is schematically shown
in Fig.~\ref{falcke-f1}. The interesting point here is that the
smallest scale in a system will be set by the size of the black
hole. Hence stellar mass black holes will be able to produce flat
spectra up to a maximum frequency $\nu_{\rm ssa,max}$ that is much
higher than the supermassive black holes in quasars -- by a factor
given by the ratio of the black hole masses which is of order
$10^{5-8}$. Therefore XRBs should be much more likely to exhibit
direct synchrotron emission in NIR, UV and even X-rays than normal
AGN.

\begin{figure}
\begin{center}
\includegraphics[width=0.8\textwidth, angle=-90]{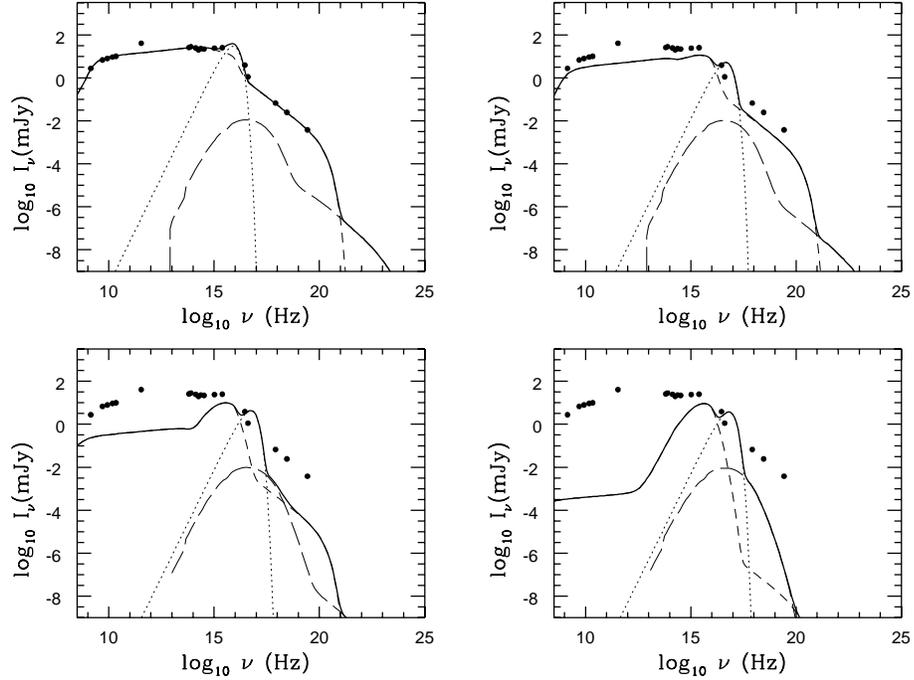}
\end{center}
\caption[]{The self-similar jet model for XRBs: The top left 
panel is the fit of a jet model to the broad-band data of XTE
J1118+480 as shown in \citeN{MarkoffFalckeFender2001} together with
black body emission from a standard optically thick disk at a large
transition radius. The subsequent panels (left to right and top to
bottom) show how the spectrum theoretically would evolve from a
low/hard-tate to a high/soft-state if one moves the transition radius
closer to the black hole. Inverse Compton cooling by disk photons
strongly quenches the hard synchrotron emission in the X-rays and
suppresses the flat radio spectrum. (From Markoff et al., in prep.)}
\label{falcke-f2}
\end{figure}

This is particularly interesting for the hard X-ray power-laws
observed in some XRBs. In a jet spectrum, beyond the turnover point of
the flat radio spectrum at $\nu>\nu_{\rm ssa,max}$, the synchrotron
emission is optically thin. The shape of the spectrum depends on the
electron distribution on the smallest scales in the jet. A thermal
distribution leads to an exponential cutoff, but a power-law
distribution which is typically observed in luminous AGN jets leads to
a hard spectrum with spectral indices ranging from $\alpha=-0.5$ to -1
(energy flux density index: $S\nu\propto\nu^{\alpha}$). The maximum
frequency of the optically thin spectrum can be found by balancing
acceleration and radiation loss times and, as
\citeN{MarkoffFalckeFender2001} (MFF01) showed, can in principle
easily reach several 100 keV fairly independent of the jet power or
shock location. MFF01 also showed that such a model can well explain
the broad-band spectrum of the X-ray binary XTE J1118+480 (see
Fig.~\ref{falcke-f2}, top left). Therefore X-ray emission from jets in
XRBs is something that has to be dealt with in understanding
the spectra of stellar mass black holes.

Of course, the situation may not always be so simple as described in
MFF01 for XTE J1118+480, a source which shows almost no sign of
reflection components from an accretion disk in the observed spectra
\cite{MillerBallantyneFabian2001}. Under different circumstances, the disk itself
should have a more direct or indirect influence on the X-ray
spectrum. One such effect is radiation cooling of highly relativistic
electrons in the jet due to photons from the accretion disk. This is
important when the accretion rate increases, the disk luminosity
increases and, in the truncated disk model, the inner edge of the
optically thick disk approaches the inner region of the jet. The
drastic increase in photon density near the black hole can then cool
most hot electrons and leave the jet very radio quiet as well as
suppress the hard X-ray power-law. This situation is shown in
Fig.~\ref{falcke-f2}, where we start with the published jet spectrum
in XTE J1118+480 and move the transition radius arbitrarily closer to
the black hole, taking radiation cooling into account. As expected,
radio and hard X-rays disappear, while the black-body emission of the
accretion disk appears in soft X-rays. There is only some jet
contribution to the EUV and soft X-ray spectrum from the jet nozzle
left.  Therefore, a change in transition radius of the accretion disk
could at least qualitatively explain the transition from low-hard to
high-soft state and explain why radio and hard X-rays seemingly
disappear while the accretion rate goes up.

Such a spectral transition would be difficult to observe in AGN due to
the much longer timescales. However, one cannot help to wonder whether
such a behavior could also explain the differences between BL Lacs and
luminous blazars (or between FR I and FR II radio galaxies/radio-loud
quasars as their respective host populations). After all, BL Lacs are
intrinsically less luminous than beamed radio-loud quasars but produce
much harder and more energetic spectra, extending up to TeV energies
with little evidence for disk emission.

\section{Microblazars as Lighthouses of the Nearby Universe}
The idea that jets in XRBs contribute to the X-ray
spectrum\footnote{In this context it is interesting to note that the
possibility of an angle-dependent jet contribution in X-rays was
already raised by \citeN{ShakuraSunyaev1973}.}  has some other
interesting consequences. If jet-emission is significant for edge-on
sources like XTE J1118+480 it will be even more important when the
source points towards the observer. In analogy to AGN, where jets
pointing toward the observer are believed to cause the strong blazar
emission from radio to gamma-rays, a microquasar pointing towards the
observer should appear as a `microblazar'
\cite{MirabelRodriguez1999}. 

This immediately raises the question whether (some of) the
ultra-luminous off-nuclear X-ray sources (ULXs), also called
Super-Eddington sources or intermediate X-ray objects (IXOs;
\citeNP{ColbertMushotzky1999}), that have been discussed at this
conference (Makishima), are in fact such microblazars. As commonly
done in unified schemes for AGN \cite{UrryPadovani1995} the population
of relativistically beamed sources is rather well defined, once one
specifies a host population and appropriate parameters for the jets --
particularly the bulk Lorentz factor $\gamma_{\rm j}$. This allows one
to check the validity of such ideas.

Along these lines \cite{KoerdingFalckeMarkoff2001} have investigated a
simple population synthesis model for XRBs and compared it
with X-ray point source populations in nearby galaxies detected in
recent Chandra observations.

\begin{figure}
\begin{center}
\includegraphics[width=0.5\textwidth, angle=-90]{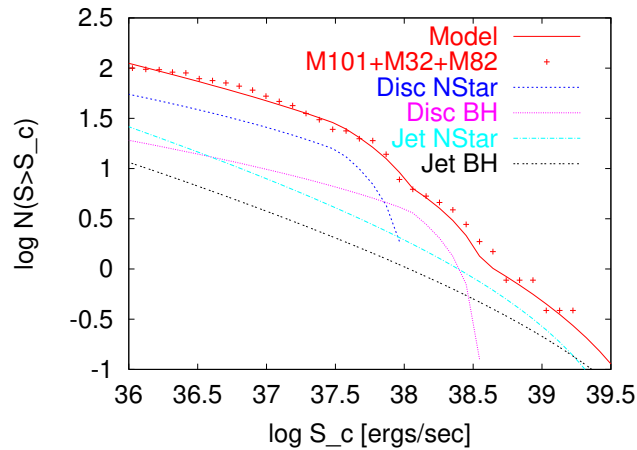}
\end{center}
\caption[]{Log N/Log S distribution of X-ray sources (dots) in three nearby
galaxies together with a simple population synthesis model (lines) for
isotropic disk and anisotropic, relativistically beamed jet emission
for XRBs in low/hard and high/soft states. (From K\"ording et al. 2001)}
\label{falcke-f3}
\end{figure}

The model consists of basically two populations of XRBs,
neutron stars and black holes, which emit X-ray emission isotropically
from an accretion disk and anisotropically from a jet due to
relativistic beaming. Below a critical accretion rate $\dot{M}_{\rm c}$
(10\% of Eddington) the disk luminosity is assumed to scale as $L_{\rm
disk}\propto \dot M^2$ (according to the ADAF paradigm) and above
$\dot{M}_{\rm c}$ as $L_{\rm disk}\propto \dot M$. The jet emits with
$L_{\rm jet}\propto \dot M^{1.4}$ (according to
\citeNP{FalckeBiermann1995}) and above $\dot{M}_{\rm c}$ as $L_{\rm
jet}\propto \dot M$ in the radiation-cooling dominated regime
mentioned above. Accretion rates ($\dot M$) are assumed to be
distributed in a power-law. Figure~\ref{falcke-f3} shows the result of
this beaming model together with the data for a $\gamma_{\rm j}=5$
jet. This shows that the high-luminosity end can indeed be explained
by microblazars without violating the low-luminosity end of the
distribution or making extreme assumptions about the properties of
jets. Hence, as blazars are the most luminous lighthouses of the
distant universe, microblazars could be the lighthouses of the local
universe.

\section{Circular Polarization and the Nature of Jets}
So far we have merely used the fact that jets exist and radiate to
discuss their observable impact. Their emission and kinetic power
depends significantly on their mass content and the electron
distribution -- specifically the distribution of hot electrons.
Recent observations of radio circular polarization in AGN, X-ray
binaries, and the Galactic Center black hole
\cite{WardleHomanOjha1998,BowerFalckeBacker1999,FenderRaynerNorris2000}
suggest that this may only be the tip of the iceberg. 

\begin{figure}
\begin{center}
\includegraphics[bb=0.7cm 7.6cm 19.9cm 24.7cm, width=0.5\textwidth]{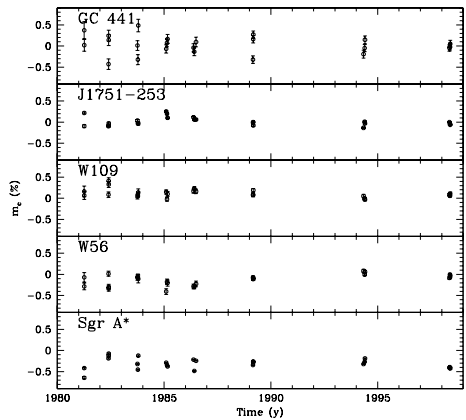}
\end{center}
\caption[]{20 Years light curve of circular polarization measurements
in Sgr A* and nearby calibrator sources with the VLA at 5
GHz. Evidently the level and sign of circular polarization remains
constant for Sgr A* over the entire period, while short-term
variability may be present. (From Bower et al., in prep.)}
\label{falcke-f4}
\end{figure}

One example of such measurements is shown in Fig.~\ref{falcke-f4},
where we plot a 20 year light curve of circular polarization in Sgr A*
\cite{BowerFalckeSault2002} -- the radio source coincident with the supermassive black hole at the
Galactic Center \cite{MeliaFalcke2001}. The source shows about 0.5\%
circular polarization with a stable sign despite strong flux
variations during this time. There is also no detectable linear
polarization. The overall spectrum of the source can be understood in
terms of a jet model \cite{FalckeMarkoff2000}.

\begin{figure}
\begin{center}
\includegraphics[bb=1.6cm 20.2cm 20.2cm 25.6cm, width=0.99\textwidth]{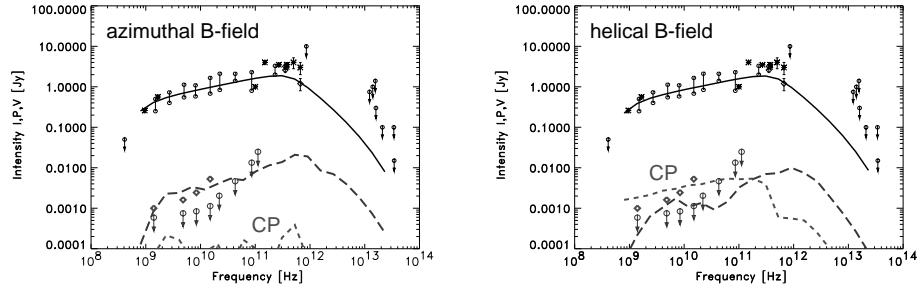}
\end{center}
\caption[]{Data and jet model for polarization in Sgr A*: small
circles -- average measured total intensity, big circles -- measured
upper limits for linear polarization, small diamonds -- measured
circular polarization, solid line -- jet model total intensity,
long-dashed line -- jet model linear polarization, short-dashed line
-- jet model circular polarization. The left panel is for a turbulent
plus a purely azimuthally ordered magnetic field, the right panel is
for a turbulent plus a helically ordered magnetic field. A pure
azimuthal (plus turbulent) magnetic field is not able to produce
significant circular polarization. (From Beckert \& Falcke, in prep.)}
\label{falcke-f5}
\end{figure}

Circular polarization can be produced through conversion by a
bi-refringent medium (such as a magnetized plasma; see
\citeNP{JonesODell1977}) where the magnetic field has a component transverse to the 
line-of-sight and the radio waves are Faraday rotated. De-polarization
of linear polarization can be obtained by random Faraday rotation in a
turbulent plasma where field components are along the
line-of-sight. Both processes are sensitive to the presence of
low-energy electrons. The ratio of linear to circular polarization in
a jet can be calculated with an appropriate radiation transfer code
for various parameters (Beckert \& Falcke, in
prep.). Fig.~\ref{falcke-f5} shows the result of such a calculation
for the case of Sgr A* and two magnetic field configurations: a
helical and a purely poloidal magnetic field on top of a turbulent
field.  Only the former can produce the observed level of circular
polarization. Since field-reversals cancel any effect of Faraday
rotation and conversion, one needs a field configuration with a
dominating component of {\it one polarity} along the line-of-sight (or
along the jet axis for moderately inclined jets).

Such a configuration is most naturally achieved by a helical magnetic
field as is presumed to exist in jets. In addition, the number of
low-energy electrons producing the conversion and de-polarization
needs to be significantly (by 2-3 orders of magnitude) higher than the
number of radiating hot electrons.  For Sgr A* this increase in
particle numbers means that the mass outflow rate and total jet power
can be orders of magnitude higher than inferred so far. The stability
of circular polarization also indicates that the polarity of the
magnetic field (the magnetic north pole) has remained constant over
the last two decades. Given the rather short accretion time scale in
this source one could speculate that this polarity is related to the
accretion of a stable large-scale magnetic field which is accreted and
expelled by a jet close to the black hole.

\section{Conclusions}
Jets are a major source of luminous emission in AGN. Our modeling of
jet spectra suggests that this is also the case for X-ray
binaries. Jets can contribute to the hard X-rays, and in microblazars
X-rays could be beamed to apparent super-Eddington luminosities. The
power of these jets could be relatively large, as indicated by the
recent detection of circular polarization in various sources which
implies a large number of low-energy electrons. The long-term
stability of the sign of circular polarization indicates that these
jets have a non-vanishing magnetic flux and stable north pole.


%

\end{document}